\documentclass[onecolumn,aps,10pt,amssymb,amsmath,amsfonts]{revtex4}
\usepackage{graphicx}
\usepackage[usenames,dvipsnames]{xcolor}
\usepackage{mathrsfs}
\usepackage{bm}
\usepackage{wasysym}    
\usepackage{color}
\def\be{\begin{equation}}
\def\ee{\end{equation}}
\def\bc{\begin{center}}
\def\ec{\end{center}}

\newcommand{\dagga}{{\phantom{\dagger}}}
\newcommand{\bea}{\begin{eqnarray}}
\newcommand{\eea}{\end{eqnarray}}

\usepackage[colorlinks,bookmarks=false,citecolor=blue,linkcolor=Green,urlcolor=NavyBlue]{hyperref}

\begin{document}
\title{Spin wave theory for 2D disordered hard-core bosons}

\author{Juan Pablo \'Alvarez Z\'u\~niga}
\author{Gabriel Lemari\'e}
\author{Nicolas Laflorencie}
\affiliation{Laboratoire de Physique Th\'eorique, Universit\'e de Toulouse, UPS, (IRSAMC), F-31062 Toulouse, France}
\begin{abstract}
A spin-wave (SW) approach for hard-core bosons is presented to treat the problem of two dimensional boson localization in a random potential. After a short review of the method to compute $1/S$-corrected observables, the case of random on-site energy is discussed. Whereas the mean-field solution does not display a Bose glass (BG) phase, $1/S$ corrections do capture BG physics. In particular, the localization of SW excitations is discussed through the inverse participation ratio.
\end{abstract}

\maketitle
\section{Introduction}
The problem of disordered superfluids (SF) and superconductors has attracted an increasing interest over the past decades~\cite{Ma85-Ma86,Fisher89, Krauth91, Wallin94, Baranger06, Prokofiev04-Pollet09,Sacepe08-11,Sanchez10,Altman10,Zoran12,Ioffe10-Feigelman10,Benfatto12,Lemarie13,mueller2013}. Numerous theoretical techniques, such as mean-field, scaling theory, renormalization group, quantum Monte Carlo simulations, and cavity mean-field have been used to investigate the localization of interacting bosons in presence of disorder. Contrary to fermions where the non-interacting case is a good starting point to understand the physics of localization~\cite{Evers08}, non-interacting bosons are pathological since all bosons will condense into the lowest single particle state. The opposite limit is achieved when a hard-core constraint is imposed, such that no more than one boson can live on each site. This hard-core condition has been shown to be physically relevant in various situations: lattice model for Helium II~\cite{Matsubara56}, preformed cooper pairs in localized superconductors~\cite{Ma85-Ma86}, spin-gapped antiferromagnets in an external field~\cite{Giamarchi08}. 

In this paper, we will discuss two dimensional hard-core bosons models using a spin wave approximation, as recently discussed for clean~\cite{Coletta12} and disordered lattices~\cite{Alvarez13}.  It is indeed quite appealing to ask whether boson localization can be captured when the first quantum corrections (namely $1/S$ corrections using linear spin wave theory) are included above the mean-field solution where Bose glass (BG) physics is absent. The Hamiltonian we will study is the following:
\be
{\cal {H}}=-t\sum_{\langle ij\rangle}\left(a_i^{\dagger}a_j^{\vphantom\dagger}+a_j^{\dagger}a_i^{\vphantom\dagger}\right)-\sum_i\mu_in_i,
\label{eq:1}
\ee
where $t$ is the hopping between neighboring sites, $\mu_i$ the (possibly random) chemical potential and $a_i^\dagger$ $(a_i)$ denotes the operator creating (destroying)
a hard-core boson at site $i$. An interesting property of hard-core bosons is that they can
be exactly mapped onto spins $\frac{1}{2}$ using the Matsuda-Matsubara mapping~\cite{Matsubara56}:
$n_i=S_i^z+1/2$, $a_i^\dagger=S_i^+$ and $a_i=S_i^-$.
The equivalent $S=\frac{1}{2}$ model is simply an XY model in a transverse magnetic field $\mu_i$:
\be
\mathcal{H}=-2t\sum_{\left\langle ij\right\rangle}{\left(S_i^x S_j^x + S_i^y S_j^y \right)}-\sum_i\mu_i\left(S_i^z+\frac{1}{2}\right).
\label{eq:2}
\ee
Based on such a mapping, a semi-classical approximation can be developed starting from the large $S$ limit
of this magnetic Hamiltonian. This approach has been developed in a series of papers \cite{Scalettar95,Murthy97,Bernardet02,Coletta12}. Below, we review the key steps of this semi-classical treatment, and we discuss the SF - BG transition for bosons in a random potential.
\section{Semi-classical treatment of hard-core bosons}
\subsection{Spin wave approximation}
Having rewritten the hard-core bosonic model \eqref{eq:1} as a spin Hamiltonian \eqref{eq:2}, one first performs a classical treatment, replacing spin operators by 3D vectors:
$\vec{S}_{i}=S \left(\sin\theta_i\cos\varphi_i,\sin\theta_i\sin\varphi_i,\cos\theta_i\right)$.
The classical energy then reads
\be
E_0=-2tS^2\sum_{\langle i j \rangle}\sin\theta_i\sin\theta_j\cos\left(\varphi_i-\varphi_j\right)-\sum_i\mu_i S\cos\theta_i.
\ee
In the absence of a twist at the boundary (useful to compute the  response~\cite{Fisher73}, see below), the energy is minimized for $\varphi_i=$ constant, and 
\be \mu_i\sin\theta_i=2tS\cos\theta_i\sum_{j{~\rm nn~}i}\sin\theta_j,\label{eq:4}\ee for all sites $i$, where $j{~\rm nn~}i$ are the nearest neighbors of $i$. In the clean case, translational invariance simplifies the problem, yielding for all sites
\be
\cos\theta=\frac{\mu}{8tS}.
\ee
For random potentials $\mu_i$, the minimization of the classical energy \eqref{eq:4} generally cannot be done analytically, except for the special bimodal case where $\mu_i=\pm W$ with probability $1/2$. Indeed, for such a disorder distribution, the classical angles satisfies
\be
\cos\theta_i=\pm\frac{W}{8tS}~~~~{\rm and}~~~~\sin\theta_i=\sqrt{1-\left(\frac{W}{8tS}\right)^2}.
\ee
For more general dense distributions of the $\mu_i$'s, for instance square box distribution ($P(\mu)=(2W)^{-1}$ if $|\mu|\le W$, and $=0$ otherwise), we have to numerically solve \eqref{eq:4} using an iterative process.
Once the classical angles are determined, we perform a rotation of the spin operators such that the new quantization axis $z'$ is aligned with the classical vector:
\begin{equation}\label{eq:rot}
 \begin{array}{lll}
  S_i^x &=& \left(\cos\theta_i\right) S_i^{x^\prime}+ \left(\sin\theta_i\right) S_i^{z^\prime} \\
  S_i^y &=& S_i^{y^\prime} \\
  S_i^z &=& \left(\cos\theta_i\right) S_i^{z^\prime}- \left(\sin\theta_i\right) S_i^{x^\prime}.
 \end{array}
\end{equation}
The new spin operators can be expressed in terms of Holstein-Primakoff bosons~\cite{Holstein40}:
\begin{equation}\label{eq:HP}
S_i^{z^\prime}=\displaystyle S-b_i^\dagger b_i^\dagga,~~
S_i^{x^\prime}=\displaystyle \frac{\sqrt{2S}}{2} (b_i^\dagga+b_i^\dagger)+\ldots,~~S_i^{y^\prime}=\displaystyle \frac{\sqrt{2S}}{2i}(b_i^\dagga-b_i^\dagger)+\ldots
\end{equation}
Combining the rotations \eqref{eq:rot} and the Holstein-Primakoff representation \eqref{eq:HP}, when only linear corrections are kept, the original hard-core bosonic Hamiltonian \eqref{eq:1} reads:
\be\label{eq:H2}
 \mathcal{H}= E_0-\sum_{\langle ij\rangle}\left[t_{ij}\left(b_i^\dagga b_j^\dagger+b_j^\dagga b_i^\dagger\right) + {\overline{t}}_{ij}\left(b_i^\dagga b_j^\dagga+b_j^\dagger b_i^\dagger\right)\right]+\sum_i\epsilon_i n_i+\cdots,
 \ee
where $E_0$ is the classical energy, $t_{ij}=tS(1+\cos\theta_i\cos\theta_j)$, ${\overline{t}}_{ij}=tS(\cos\theta_i\cos\theta_j-1)$, $\epsilon_i=\mu_i\cos\theta_i+2tS\sin\theta_i\sum_{j~{\rm nn}~i}\sin\theta_j$, and the ellipses denotes higher order terms.
Hamiltonian \eqref{eq:H2} is then straigthforwardly diagonalized using a generalized Bogoliubov transformation.
%
New bosonic operators $\alpha_i$ diagonalize the quadratic Hamiltonian, such that~\eqref{eq:H2} simplifies to
\be
{\cal H}=E_0+2\sum_{p=1}^{N}\Omega_p\left(\frac{1}{2}+\alpha_{p}^{\dagger}\alpha_{p}^{\dagga}\right)-\sum_{i=1}^{N}\frac{\epsilon_i}{2}.
\ee
The $1/S$-corrected ground-state energy can be easily evaluated since the new ground-state corresponds to the vacuum of Bogoliubov quasi-particles $\langle\alpha_p^\dagger\alpha_p^\dagga\rangle=0~\forall p$, yielding
\be
E_{{1}/{S}}=E_0+\sum_p\Omega_p-\sum_i\frac{\epsilon_i}{2}.
\ee
\subsection{$1/S$ computation of physical observables}
The question of the correct determination of $1/S$-corrected expectation value of a physical observable $\hat {O}$ has been clarified recently in Ref.~\cite{Coletta12}. There, we have shown that if it is possible to express ${\hat O}$ as a $n^{\rm th}$ derivative of the Hamitonian $\cal H$ with respect to an external field $\Gamma\to 0$, then it is very convenient to compute $1/S$ corrections (and higher order corrections as well) using
\be
\langle {\hat{O}}\rangle_{1/S}=\frac{\partial^{n}E_{1/S}}{\partial \Gamma^n}\Bigr|_{\Gamma=0}.
\ee
\begin{figure}
\includegraphics[height=.25\textheight]{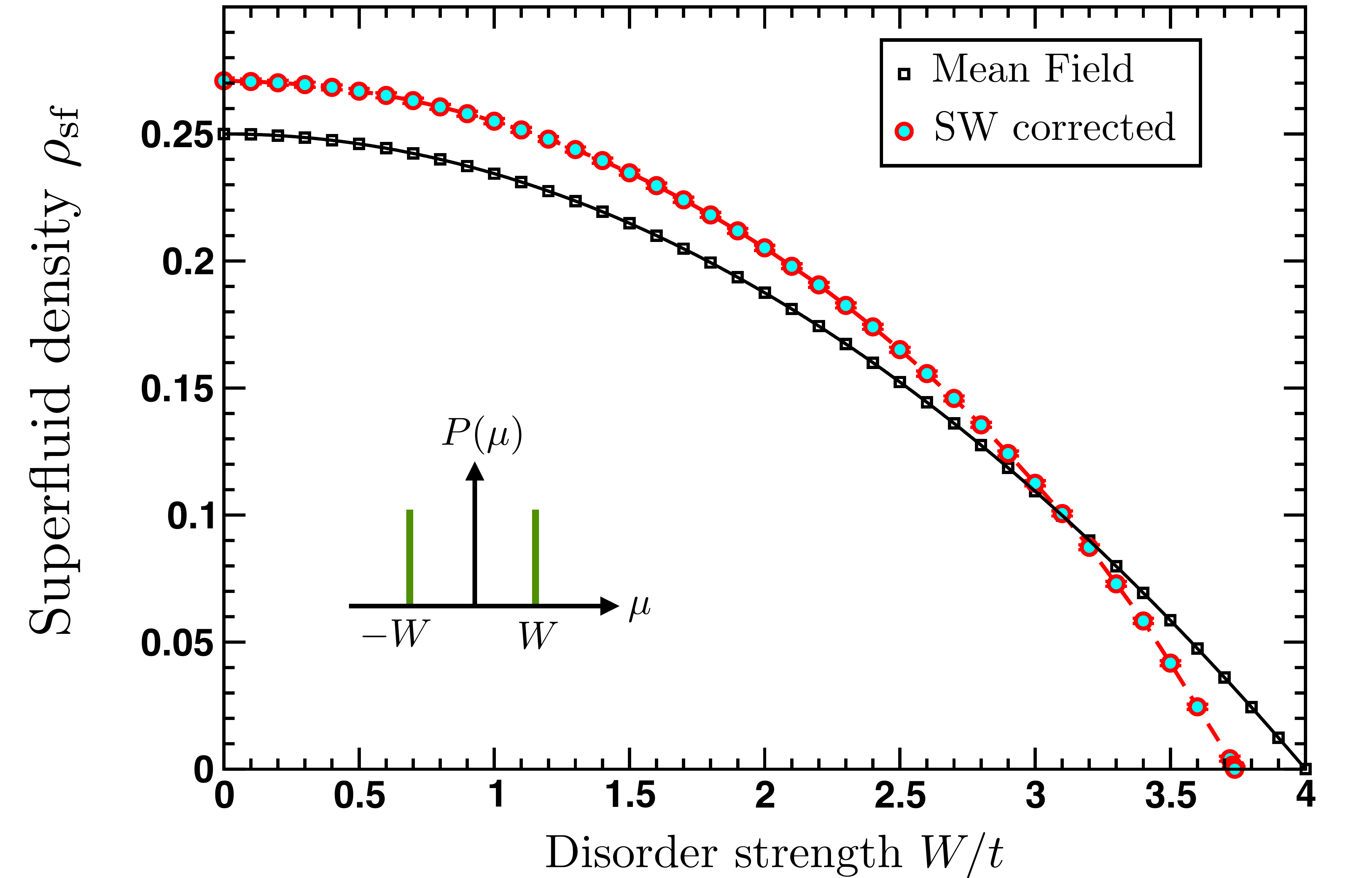}
\caption{SF density $\rho_{\rm sf}$ of two-dimensional hard-core bosons plotted against disorder strength $W/t$ for bimodal disorder, i.e. $\mu_i=\pm W$, for mean-field and $1/S$ spin-wave approximations. While at the mean-field level there is no intervening BG between SF at $W/t\le 4$ and gapped insulator for $W/t>4$, $1/S$ corrections can stabilize a BG for $3.74\le W/t\le 4$. SW results are thermodynamic limit extrapolations (see Ref.~\cite{Alvarez13}) and disorder average has been performed over several hundreds of independent samples. The inset shows the bimodal distribution of chemical potentials $P(\mu)$.}
\label{fig:1}
\end{figure}

Let's illustrate this on a simple example, assuming we want to compute the $T=0$ SF density of our original hard-core bosons $\rho_{\rm sf}$.
The SF density can be obtained by imposing a phase gradient
$\Phi_{{i}+{\vec e}}-\Phi_{i}=\varphi$ to the system (${\vec e}$ being the unit vector along one axis $x$ or $y$ of the lattice). 
Following Fisher, Barber and Jasnow~\cite{Fisher73}, the density of kinetic energy of a SF of density $\rho_{\rm sf}$ flowing at velocity $v_{\rm sf}=\left(\hbar/m^*\right)\varphi$ in one direction is given by 
\be
E(\varphi)-E(0)=\frac{1}{2}m^*\rho_{\rm sf}\left({v_{\rm sf}}\right)^2
\ee
thus yielding a SF density
\be
\rho_{\rm sf}=\frac{m^*}{\hbar^2}\frac{\partial^2 E(\varphi)}{\partial \varphi^2}\Bigr|_{\varphi=0},
\ee
where the effective mass is given by
$2m^*/\hbar^2=1/(2t)$.
In order to evaluate the SW corrections to the SF density, one needs to compute the SW-corrected energy with a small twist angle at the boundaries. While in the disorder-free situation, the global twist $\Phi$ will be uniformly distributed along the $L$ bonds $\varphi=\Phi/L$, this will not be necessary the case for general non-translationally invariant problems~\cite{Paramekanti98,Seibold12}. We therefore introduce different local twist angles directly on the bosonic operators: $b^{\dagger}_{i}\to b^{\dagger}_{i}~{\rm{e}}^{i\Phi_{i}}$ and $b_{i}^{\dagga} \to b_{i}^{\vphantom{\dagger}}~{\rm{e}}^{-i\Phi_{i}}$, leading to new local rotations of spin operators Eq.~\eqref{eq:rot}:
\bea
S^x_{i}&=&\left(\cos\theta_i S^u_{i} +\sin\theta_i S^w_{i}\right)\cos \Phi_{i}-S^v_{i}\sin \Phi_{i}\nonumber\\
S^y_{i}&=&\left(\cos\theta_i S^u_{i} +\sin\theta_i S^w_{i}\right)\sin \Phi_{i}+S^v_{i}\cos \Phi_{i}\nonumber\\
S^z_{i}&=&-\sin\theta_i S^u_{i} +\cos\theta_i S^w_{i}.
\eea
Therefore, in the presence of a global twist angle $\Phi$, the $1/S$ SW Hamiltonian reads
\be\label{eq:H2phi}
 \mathcal{H}(\Phi)= E_0(\Phi)-\sum_{\langle ij\rangle}\left[t_{ij}(\varphi_i,\varphi_j)\left(b_i^\dagga b_j^\dagger+b_j^\dagga b_i^\dagger\right) + {\overline{t}}_{ij}(\varphi_i,\varphi_j)\left(b_i^\dagga b_j^\dagga+b_j^\dagger b_i^\dagger\right)\right]+\sum_i\epsilon_i(\varphi_i,\varphi_{j{\rm~nn}~i}) n_i,
 \ee
where $E_0(\Phi)$ is the mean-field energy in the presence of $\Phi$, $t_{ij}(\varphi_i,\varphi_j)=tS\cos(\varphi_i-\varphi_j)(1+\cos\theta_i\cos\theta_j)$, ${\overline{t}}_{ij}(\varphi_i,\varphi_j)=tS\cos(\varphi_i-\varphi_j)(\cos\theta_i\cos\theta_j-1)$, and $\epsilon_i(\varphi_i,\varphi_{j{\rm~nn}~i})=\mu_i\cos\theta_i+2tS\sin\theta_i\sum_{j~{\rm nn}~i}\sin\theta_j\cos(\varphi_i-\varphi_j)$.

The $1/S$-corrected SF density 
\be
\rho_{{\rm sf},(1/S)}=\frac{1}{4t}\frac{\partial^2 E_{(1/S)}(\Phi)}{\partial \varphi^2}\Bigr|_{\Phi=0},
\ee
only requires to compute the $1/S$ correction to the energy in the presence of $\Phi$. Mean-field and SW results for $\rho_{\rm sf}$ are displayed in Fig.~\ref{fig:1} versus the disorder strength $W/t$ for the case of bimodal disorder $\mu_i=\pm W$ with probability 1/2. At the mean-field level, the SF density is simply given by $\frac{1}{4} \sin^2\theta=1/4-(W/8t)^2$ (black curve in Fig.~\ref{fig:1}) for $W/t\le 4$, and zero for $W/t>4$ where the system becomes a gapped insulator. In the SF regime, SW fluctuations lead to interesting corrections (red curve in Fig.~\ref{fig:1}): for small disorder, superfluidity is enhanced, as compared to mean-field~\cite{Note1}, whereas for stronger disorder $W/t>3$ quantum fluctuations and disorder start to cooperate to destroy superfludity which is found to vanish at $W_c/t\simeq 3.74$, before the mean-field transitioin point. Therefore a small but finite Bose-glass window, intervening between  and gapped insulator, is found at $1/S$ order, while absent in mean-field. %

\section{Excitation spectrum}
\begin{figure}[!h]
\includegraphics[height=.3\textheight]{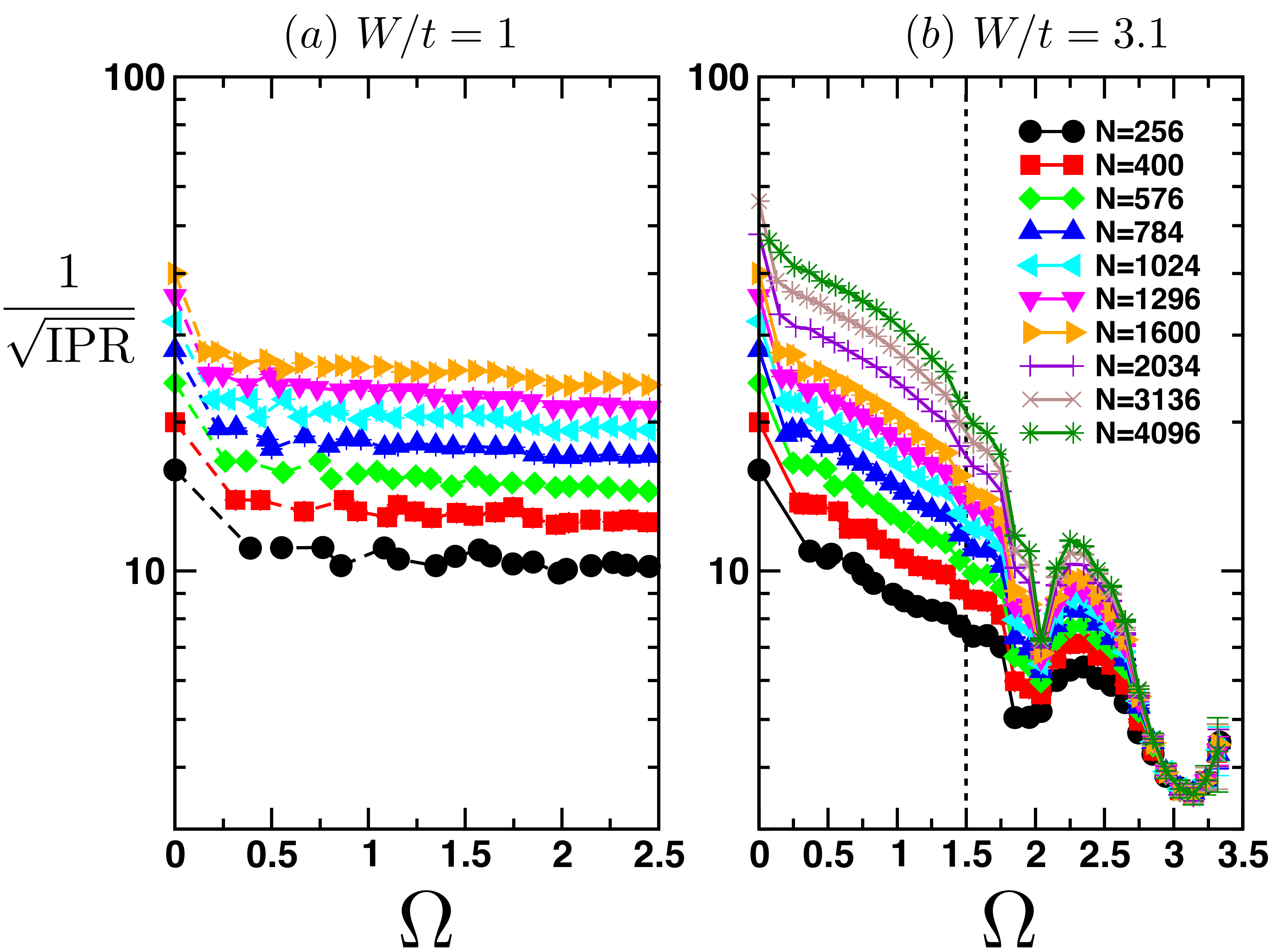}
\label{fig:2}
\caption{Representative results for IPRs in the  phase. (a) For weak disorder $W/t=1$, $1/\sqrt{\rm IPR}$ keeps increasing with $N$ for all frequencies, meaning that all modes are delocalized. (b) For stronger disorder (but still in the SF phase), a qualitative difference is visible in the spectrum between low and high energy states. A quantitative analysis is proposed in Fig.~\ref{fig:3} where a localization transition is observed at $\Omega_c\simeq 1.5$, as shown here by the dashed vertical line.}
\end{figure}

\begin{figure}
\includegraphics[height=.25\textheight]{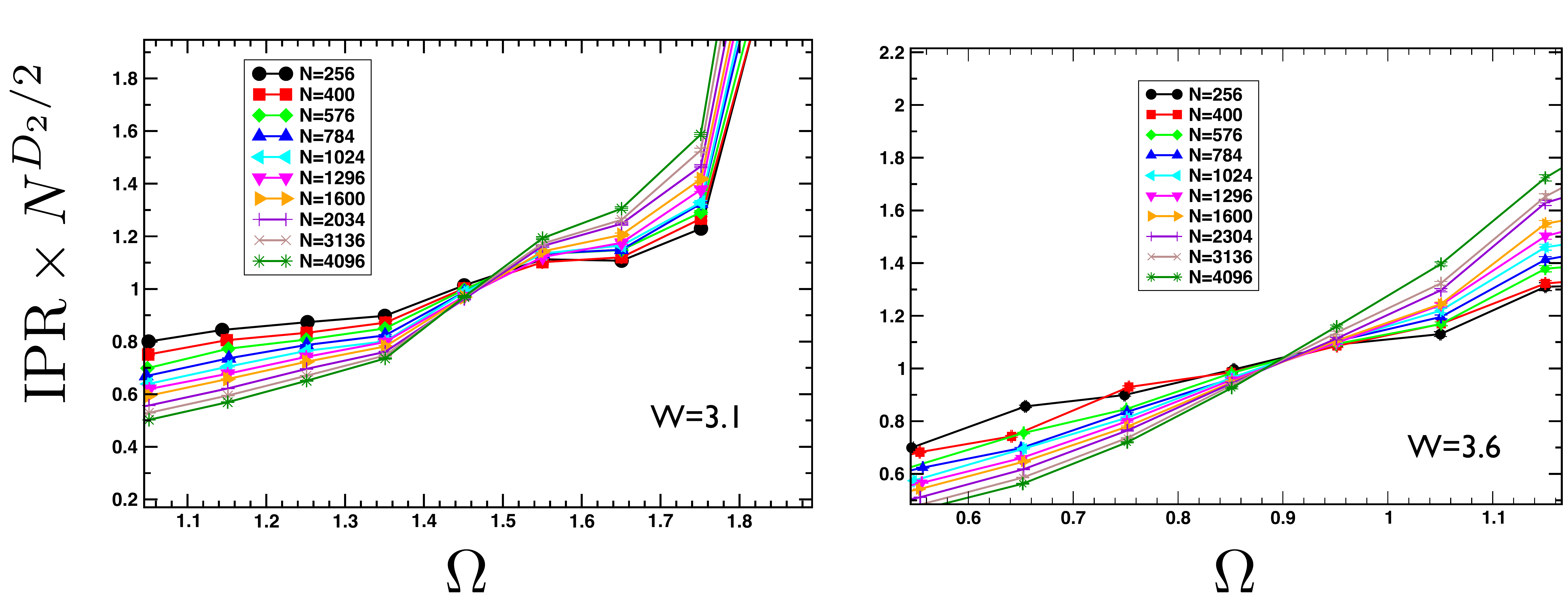}
\caption{Best crossing of IPR$\times N^{D_2/2}$ obtained with $N=256,\cdots, 4096$ at a mobility edge $\Omega_c\simeq 1.5$ for $W/t=3.1$ (left) and  $\Omega_c\simeq 0.9$ for $W/t=3.6$ (right), both using the fractal dimension $D_2=1.48$. Results obtained for bimodal disordered hard-core bosons on the square lattice.}
\label{fig:3}
\end{figure}

An interesting question concerns the spin-wave (bosonic) excitation spectrum in the presence of disorder, a topic only addressed in a few works~\cite{Mucciolo04,Wessel05,Monthus10-Amir13,Vojta13}. Here we want to address this question using the inverse participation ratio (IPR) for single particle Bogoliubov excitations, usually defined as IPR$_p=\sum_j|\psi_j^p|^4$ for normalized states $|\Psi_p\rangle=\sum_j\psi_j^p|j\rangle$, with $j=1,\cdots,N=L^d$ the lattice sites . For delocalized states, IPR$_p\sim 1/L^d$, it saturates to a finite value for localized states: IPR$_p\sim (1/\xi_p)^d$, and an anomalous scaling is expected at the localization-delocalization transition: IPR$_p\sim 1/L^{D_2}$, with $D_2\le 2$ the fractal dimension.
Starting from the bosonic Bogoliubov transformation which diagonalized the quadratic spin-wave Hamiltonian
$b_i^\dagger=\sum_p(u_{ip}\alpha_p^\dagga+v_{ip}\alpha_p^\dagger)$,
we use the following definition for the IPRs~\cite{Mucciolo04,Wessel05}
\be
{\rm{IPR}}_p=\frac{\sum_{i}|v_{ip}|^4}{\left(\sum_{i}|v_{ip}|^2\right)^2}
\ee
for each eigenmode $p$. Then, in order to make a frequency-dependent study, we average over finite slices of frequencies centered around $\Omega$:
\be
{\rm IPR}(\Omega)=\frac{\sum_p\Theta(\Omega_p,\Omega\pm\delta\Omega){\rm IPR}_p}{\sum_p \Theta(\Omega_p,\Omega\pm\delta\Omega)},
\ee
where $\Theta(\Omega_p,\Omega\pm\delta\Omega)=1$ if $\Omega-\delta\Omega\le\Omega_p\le\Omega+\delta\Omega$, and 0 otherwise, with $\delta\Omega/v_0=1/20$ in the following. In Fig.~\ref{fig:2} we present two representative results in the SF phase: (a) small disorder $W/t=1$ where $1/\sqrt{\rm IPR}$ keeps increasing with $N$ for all frequencies, signaling that all eigenmodes are extended, in agreement with what is expected for Goldstone modes~\cite{Gurarie03}; (b) strong disorder $W/t=3.1$ where a qualitative difference is clearly visible between low and high energy states. Indeed, for such a disorder, whereas low energy modes are clearly delocalized, again in agreement with Ref.~\cite{Gurarie03}, a transition occurs at finite frequency in the spectrum. In order to precisely locate this transition, we exploit the fractal scaling at the transition $IPR\sim N^{-D_2/2}$. Plotting in Fig.~\ref{fig:3} IPR$\times N^{D_2/2}$ for
$N=256,\cdots, 4096$, we get the best crossing using $D_2=1.48$ which works not only for this particular value of the disorder $W/t=3.1$ but also for $W/t=3.6$, as shown in the right panel of Fig.~\ref{fig:3}, and for other disorders~\cite{Alvarez13}. These crossings signal a mobility edge at finite frequency $\Omega_c$, separating delocalized low energy excitations from localized ones at higher energies. As discussed in Refs.~\cite{Alvarez13,alvarez-2013}, this mobility edge is expected to vanish in the BG phase where all excited modes are localized.
\section{Conclusion}
To conclude, we have shown that $1/S$ spin wave corrections provide very interesting information about the superfluid - Bose glass transition for two-dimensional hard-core bosons in a random potential. For sufficiently strong randomness, quantum fluctuations cooperate with disorder such that the superfluid density vanishes at a critical point, leaving room for a stable gapless localized phase (the Bose glass) before entering in a gapped insulator, a phenomenology absent from a classical (mean-field) treatment. The study of the excitation spectrum above the superfluid ground-state gives non-trivial results, namely the existence of a mobility edge at finite frequency, separating delocalized modes at low energy from localized ones at higher energy.

\bibliographystyle{aipproc}

\end{document}